\documentclass[12pt]{article}
\usepackage{amssymb}
\begin{document} 

\title{Vacuum fluctuations of the supersymmetric field in curved background}
\author{
Neven Bili\'c\thanks{Electronic mail:
bilic@thphys.irb.hr},
Silvije Domazet\thanks{Electronic mail:
sdomazet@irb.hr}, 
Branko Guberina\thanks{Electronic mail:
guberina@thphys.irb.hr}
 \\
Rudjer Bo\v skovi\'c Institute,
POB 180, HR-10002 Zagreb, Croatia 
}

\maketitle

\begin{abstract}
We study a supersymmetric model in curved background spacetime.
We calculate the effective action and the vacuum expectation value of the energy momentum tensor
 using a covariant regularization procedure.
A soft supersymmetry breaking induces a nonzero contribution to the vacuum energy density and pressure.
Assuming the  presence of a cosmic fluid in addition to the vacuum fluctuations of the supersymmetric field
an effective equation of state is derived in a self-consistent approach at one loop order.
The net effect of the vacuum fluctuations of the supersymmetric fields in the  leading adiabatic order 
is a renormalization of the Newton and cosmological constants.
\end{abstract}

\maketitle


\section{Introduction}

Observational evidence for an
  accelerating expansion \cite{perlmutter,bennett,spergel}
implies that
 the vacuum energy density
dominates
the total energy density today.
The vacuum energy or cosmological constant (CC)
which was introduced {\it ad-hoc} in the Einstein-Hilbert action
is related to the vacuum energy density of matter fields.
The main CC problem is that the 
 vacuum energy contribution estimated in quantum field theory
is much larger than the observed value.
In general, the result depends quartically on a  quantity
$\Lambda_{\rm cut}$ that represents the UV momentum cutoff, i.e., 
 the vacuum energy density  goes as 
$\Lambda_{\rm cut}^4$.
If $\Lambda_{\rm cut}$ is of the order of the Planck mass, 
the estimated vacuum energy density 
is by about 120 orders of magnitude larger than the value required by
astrophysical and cosmological observations \cite{weinberg1}.
This term, if really there, would be disastrous  even if the cutoff were in the mass range of 
the standard model of particle physics.

The above mentioned estimate was based on field theory in flat spacetime
in which one can easily solve the problem by redefining the vacuum energy.
Since the energy is defined up to an arbitrary additive  constant
one can subtract the divergent contributions and make the vacuum energy exactly zero
\footnote{ Unfortunately, 
the cancellation of the divergent contributions does not solve the cosmological constant problem.
The remaining finite contributions of the type $m^4$ in a realistic field theory are still much larger then
the observed vacuum density. Besides,  all such contributions coming from different
sectors of the standard model of particle physics  should somehow  conspire in order to
reproduce the observe tiny value of the vacuum energy density. 
Extreme fine tuning is needed for this to happen.}.
However , in curved spacetime this procedure cannot be performed because
the energy is a source of the gravitational field and
adding or subtracting (even  constant) energy changes the spacetime
geometry.
In spite of that, the quantum field theory in curved spacetime is  renormalizable 
provided higher derivative terms in addition to the Einstein-Hilbert term with a cosmological constant
are introduced at the classical level.
In what follows we use the low energy formulation of quantum field theory in curved spacetime 
\cite{birrell,buchbinder}. The theory is described by the gauged matter Lagrangian in addition to
the Einstein-Hilbert action.

Phenomenologically, it is   desirable to have the vacuum energy exactly zero or comparable to
the tiny value of the critical density of the universe.
To achieve that,
one must either  kill the flat spacetime contribution
simply by {\it fiat}  or  invent a symmetry principle that forbids a nonzero vacuum energy.
Such a principle is indeed provided by supersymmetry \cite{wess}.
In field theory with exact supersymmetry,  the
contributions  of fermions and bosons to  vacuum energy precisely
cancel \cite{weinberg2}.

In references \cite{bilic,bilic2} a  residual quadratic contribution of the
form $H^2 \Lambda_{\rm cut}^2$ has been found after canceling the flat spacetime 
 parts in the SUSY limit.
Such a contribution is phenomenologically
acceptable owing to the fact that the present value of the CC
density is of the same order, if $\Lambda_{\rm cut}$ is taken to be of the order of the Planck mass.
Such quadratic contribution was also found in recent papers \cite{maggiore,sloth}
and in some earlier papers \cite{sola,stefancic} in a different context.
In particular, the work \cite{maggiore} presents a similar calculation
of the zero-point energy using only a massless boson field and obtains two types of contributions: 
the quartic type
$\Lambda_{\rm cut}^4$ and the quadratic part $H^2 \Lambda_{\rm cut}^2$. 
Then, the quartic contributions to  CC was 
canceled
on the basis of the procedure used previously in
the literature with the so-called ADM mass.
In \cite{bilic,bilic2} it has been demonstrated that
in a supersymmetric world a cancellation 
by fiat
is unnecessary because  the cancellation 
between bosons and fermions of all (not only quartically divergent) 
flat-spacetime contributions is naturally provided by supersymmetry.
However, in real world where SUSY is broken this cancellation will affect only the
quartic contribution.
If  SUSY is broken at scales $m_{\rm SUSY}$ the dominant flat-spacetime contributions will be
the quadratically divergent contribution of the type $m_{\rm SUSY}^2\Lambda_{\rm cut}^2$ and
the mass terms of the type $m_{\rm SUSY}^4$. These flat-spacetime contributions  
will of course be present in nonflat spacetime too.

Another important point of both papers \cite{bilic2} and \cite{maggiore} is 
that the vacuum fluctuations cannot  be interpreted as a part of CC because
the vacuum fluctuations do not yield the equation of state
$p=-\rho$, as a consequence of the energy
momentum tensor not having a CC form.
 This behavior was already observed in flat space time if a three-dimensional
cutoff regularization was employed \cite{ossola,akhmedov,andrianov}.
 A possible way out  has recently been suggested by Maggiore et al
\cite{maggiore2,maggiore3}.
If the regularization scheme breaks general
covariance, one must also allow for noncovariant counterterms, and these
can be chosen so to obtain fully covariant results for the renormalized 
energy density and pressure.
For example, since the quartic term
in the bare energy density and pressure does not satisfy
$p = -\rho$ (the vacuum fluctuations of a  minimally coupled scalar field gives
 $p = \rho/3$), one may introduce  appropriate
 noncovariant counterterms so that the renormalized quantities satisfy
$p_{\rm ren} =- \rho_{\rm ren}$. The explicit form of these counterterms, and a 
detailed discussion of the above
issues, has been given in \cite{maggiore3} (elaborating on
results of \cite{maggiore2}).

Using a three-dimensional cutoff $\Lambda_{\rm cut}$ procedure 
is perhaps not the most elegant way to regularize a quantum field
theory. First, the three-dimensional cutoff
violates Lorentz invariance and, as a result, the flat space time terms in the energy
density and  pressure  do not satisfy the CC equation of state.
 This is not a problem for a model with an exact supersymmetry
since these terms cancel anyway.
However, one cannot be sure that the residual EOS coming from the 
dominant contribution of the type
$H^2 \Lambda_{\rm cut}^2$ is robust,
i.e. one does not know how much it is scheme dependent.
Therefore, it would be desirable to use an explicitly covariant scheme
 to regularize the integrals.
A covariant regularization in flat space time should
yield the vacuum energy momentum tensor of the form
$T_{\mu\nu}=\rho \eta_{\mu\nu}$. Naively, in curved spacetime one would  
generalize this to the CC form $T_{\mu\nu}=\rho g_{\mu\nu}$.
However, since a curved geometry involves
the Riemann tensor and its covariant derivatives 
we may  expect the energy momentum tensor at linear curvature order to be
of the form $T_{\mu\nu}=(\alpha+\beta R ) g_{\mu\nu}+\gamma R_{\mu\nu}$ 
where $\alpha$,  $\beta$, and $\gamma$ are constants that do not depend on curvature.

The main motivation for this paper is to investigate the fate of vacuum energy
when an unbroken supersymmetric model 
is embedded in a general curved spacetime.
We propose an approach based on the effective action.
We calculate one loop contributions to the
effective and we regularize divergences using a covariant UV cutoff.
Our approach is similar to Sobreira et al \cite{sobreira}
who calculated the scalar and fermion one loop contributions to
the effective potential.
From the effective action we derive the effects
of the vacuum energy fluctuations of the supersymmetric fields on the 
expansion of the universe, specifically on the cosmological constant.
With a cutoff scale of the order of the Planck mass $m_{\rm Pl}$
the contribution of the supersymmetric field fluctuations
is found to be of the same order of magnitude as that of the cosmic fluid.
For example, in the case of CC, i.e., in the de Sitter background with the expansion parameter $H$,  
the contribution is of the order $H^2 m_{\rm Pl}^2$ no fine-tuning is needed.

Unlike in flat spacetime, the vacuum energy density turns out to be nonzero depending 
on background metric.
This type of ``soft'' supersymmetry breaking is similar to the supersymmetry breaking at finite temperature 
where the Fermi-Bose degeneracy is lifted by quantum statistics (\cite{Kratzert:2003cr} and references therein).
In addition to the supersymmetric field we assume the presence of  a cosmic fluid obeying
the equation of state of the general form $p=p(\rho)$  so that that the global geometry is 
determined by
a combined effect of both the cosmic fluid 
 and vacuum fluctuations of the supersymmetric field.


The remainder of the paper is organized as follows. 
In section \ref{model} we introduce a supersymmetric model in an arbitrary curved background.
The  calculations of the effective action at one loop order  are presented in section \ref{calculations}.
In section \ref{energy} we derive the vacuum expectation value of
 the energy momentum tensor and discuss the effective equation of state.
Concluding remarks are given 
in section \ref{conclude}.

\section{The model}
\label{model}
We consider a Wess-Zumino supersymmetric model with $N$ species and calculate the energy density of vacuum fluctuations
in arbitrary curved background. Our notations follow \cite{bilic2}.
In general, the supersymmetric Lagrangian 
for $N$ chiral superfields has the form \cite{bailin}
\begin{equation}
{\cal L}_N =  \sum_i \Phi_i^\dag \Phi_i |_D +W(\Phi)|_F + \rm h.c.\, ,
\label{eq001}
\end{equation}
where the index $i$ distinguishes the various left chiral superfields $\Phi_i$
The quantity  $W(\Phi)$ denotes the superpotential for which we take
\begin{equation}
W(\Phi) =  \sum_i \left(\frac{m_i}{2}  \Phi_i^2+ \frac{\lambda}{3}\Phi_i^3 \right) .
\label{eq002}
\end{equation}
so that the Lagrangian (\ref{eq001}) is just a sum of $N$
chiral Lagrangians ${\cal L}_i$ of each species. 
For simplicity, from now on we suppress the dependence on  $i$.
Assuming a curved background 
spacetime geometry with metric $g_{\mu\nu}$ and
 eliminating auxiliary fields by equations of motion, the Lagrangian
for each species
takes the form 
\begin{eqnarray}
{\cal L}& =& g^{\mu\nu} \phi_{,\mu}^\dag \phi_{,\nu} - V(\phi)
+ \frac{i}{4}\left(\bar{\Psi}\tilde{\gamma}^\mu\Psi_{;\mu}-
\bar{\Psi}_{;\mu}\tilde{\gamma}^\mu \Psi\right)-
\frac{1}{2} m\bar{\Psi}\Psi
\nonumber
\\
&&-\frac{\lambda}{2}
\bar{\Psi}(1-\gamma_5)\Psi\phi
-\frac{\lambda}{2}
\bar{\Psi}(1+\gamma_5)\Psi\phi^\dag ,
\label{eq003}
\end{eqnarray}
with
\begin{equation}
V(\phi)=
|m\phi+\lambda\phi^2 |^2
+\xi R |\phi|^2.
\label{eq103}
\end{equation}
Here, $\phi$ and $\Psi$ are the complex scalar and the Majorana spinor fields, respectively,
and $\tilde{\gamma}^\mu$ are the curved spacetime gamma matrices \cite{birrell}.
For completeness, we included in the scalar field potential (\ref{eq103})
the nonminimal coupling term $\xi R |\phi|^2 $ of the scalar field to the  
scalar curvature $R$. The term is needed for renormalization because, even if set $\xi=0$
in (\ref{eq003}), the loop corrections would generate in the effective action a term of this type
\cite{birrell,parker2}.

In the chiral  ($m\rightarrow 0$) limit, the Lagrangian (\ref{eq003}) becomes
 invariant under the chiral U(1) transformation
\begin{equation}
\phi \rightarrow  e^{i2\alpha}\phi ,
\label{eq1003}
\end{equation}
\begin{equation}
(1-\gamma_5)\Psi \rightarrow  e^{-i\alpha}(1-\gamma_5)\Psi ,
\label{eq1004}
\end{equation}
\begin{equation}
(1+\gamma_5)\Psi \rightarrow  e^{i\alpha}(1+\gamma_5)\Psi .
\label{eq1005}
\end{equation}
This invariance reflects the $R$-invariance of the cubic superpotential 
\cite{wess2}.
The action  
 may be written as 
\begin{equation}
S= \int d^4x \sqrt{-g}( {\cal L}_{\rm B} + {\cal L}_{\rm F}) ,
\label{eq004}
\end{equation}
where ${\cal L}_{\rm B}$ and ${\cal L}_{\rm F}$ are the boson and fermion Lagrangians, respectively.
Using
\begin{equation}
\phi=\frac{1}{\sqrt{2}}(\sigma+i\pi),
\label{eq0004}
\end{equation}
the Lagrangian for a complex scalar field $\phi$ may be expressed as a
Lagrangians for two real fields $\sigma$ and $\pi$.
Then, the potential (\ref{eq103}) becomes
\begin{equation}
V(\sigma,\pi) =
\frac{m^2+\xi R}{2} (\sigma^2+\pi^2)+
\frac{\lambda^2}{4}(\sigma^2+\pi^2)^2
+\frac{m\lambda}{\sqrt{2}}\sigma (\sigma^2+\pi^2).
\label{eq1105}
\end{equation}
Variation of (\ref{eq004}) with respect to $\bar{\Psi}$ yields the Dirac equation
in curved spacetime:
\begin{equation}
i\tilde{\gamma}^\mu\Psi_{;\mu}
- (m+\sqrt{2}\lambda\sigma-i\sqrt{2}\lambda\pi\gamma_5)\Psi =0.
\label{eq106}
\end{equation}

\section{Effective action}
\label{calculations}
In this section we derive the effective action 
in a general curved background defined by a metric $g_{\mu\nu}$ and the corresponding 
curvature tensor $R_{\mu\nu}$.
We use  the method described in \cite{parker2,parker} based on the calculation of the Feynman propagator
 at one loop order.
We introduce the background fields $\bar{\sigma}$ and $\bar{\pi}$
and redefine the fields 
\begin{equation}
\sigma \rightarrow \bar{\sigma}+ \sigma ;
\hspace{1cm}
\pi \rightarrow \bar{\pi}+ \pi .
\label{eq1010}
\end{equation}
The effective action at one loop order is given by \cite{parker2}
\begin{equation}
\Gamma[\bar{\sigma},\bar{\pi}]=S^{(0)}[\bar{\sigma},\bar{\pi}] 
-i \ln   \int [d\sigma,d\pi,d\Psi]
\exp (iS^{(2)}[\bar{\sigma},\bar{\pi},\sigma,\pi,\Psi]),
\label{eq0010}
\end{equation}
where $[d\sigma,d\pi,d\Psi] $ denotes the measure of the functional integral,
$S^{(0)}$ is the classical part of the action ({\ref{eq004})
and $S^{(2)}$ is the part of the action  which is quadratic
in quantum fields.
The classical part is just the boson part of the action (\ref{eq004})
in which the fields $\sigma$ and $\pi$ are replaced by $\bar{\sigma}$ and $\bar{\pi}$.

For the quadratic part we find
\begin{equation}
S^{(2)}= \int d^4x \sqrt{-g}( {\cal L}^{(2)}_{\rm B} + {\cal L}^{(2)}_{\rm F}) ,
\label{eq0104}
\end{equation}
where
\begin{equation}
{\cal L}^{(2)}_{\rm B} =  \frac{1}{2}  g^{\mu\nu} \sigma_{,\mu} \sigma_{,\nu}
+\frac{1}{2}  g^{\mu\nu}\pi_{,\mu} \pi_{,\nu}
-\frac{m_\sigma^2}{2} \sigma^2
-\frac{m_\pi^2}{2}\pi^2
-\frac{\xi}{2}R (\sigma^2+\pi^2)
\label{eq0105}
\end{equation}
and
\begin{equation}
{\cal L}^{(2)}_{\rm F} = 
\frac{i}{4}\left(\bar{\Psi}\tilde{\gamma}^\mu\Psi_{;\mu}-
\bar{\Psi}_{;\mu}\tilde{\gamma}^\mu \Psi\right)-
\frac{1}{2} m_{\rm F}\bar{\Psi}\Psi 
+\frac{i\lambda\bar{\pi}}{\sqrt{2}}\bar{\Psi}\gamma_5\Psi .
\label{eq0106}
\end{equation}
The effective boson masses $m_\sigma^2$ and $m_\pi^2$,  are the coefficients of the
diagonalized quadratic form in $\sigma$ and $\pi$, and $m_{\rm F}$ is the effective fermion mass . We find
\begin{equation}
m_\sigma^2=a+b; \hspace{1cm} m_\pi^2=a-b; \hspace{1cm}
m_{\rm F}=m+\sqrt{2}\lambda\bar{\sigma} ,
\label{eq201}
\end{equation}
where
\begin{equation}
a=m^2+2\lambda^2(\bar{\sigma}^2+\bar{\pi}^2)+2\sqrt{2}m\lambda\bar{\sigma} 
\label{eq202}
\end{equation}
and
\begin{equation}
b=\lambda\sqrt{(\bar{\sigma}^2+\bar{\pi}^2)\left[(\lambda\bar{\sigma} 
+\sqrt{2}m)^2 +\lambda^2\bar{\pi}^2\right]} .
\label{eq203}
\end{equation}
In the following we calculate separately the contributions of the fermion and scalar fields.
\subsection{Scalar fields}
\label{scalar}

Following Parker and Toms \cite{parker2} the contribution of each scalar field
 may be calculated using the expression
\begin{equation}\label{eq0014}
\Gamma_s=-\frac{i}{2}\int d^4x \sqrt{-g} \int^{m_s^2} d(m^2) \Delta(x,x),
\end{equation}
where the subscript $s$ stands for either $\sigma$ or $\pi$. 
The renormalized Feynman propagator
\begin{equation}
\Delta(x,x')=-i\langle \varphi(x)\varphi(x') \rangle
\label{eq0012}
\end{equation}
satisfies the equation
\begin{equation}
(-\Box -m^2 -\xi R)\Delta(x,x')=(-g(x))^{-1/2}\delta(x-x'),
\label{eq1012}
\end{equation}
where $\Box=g^{\mu\nu}\nabla_\mu\nabla_\nu$.
In the coincidence limit, $x \rightarrow x'$,
\begin{equation}
\Delta (x,x) = \int \frac{{d^d k}}{(2\pi)^d} \left[ 1 + \sum_{j=1}^{\infty} \overline f_j  (-\frac{\partial }{\partial m^2})^j \right] \left[ k^2 -m^2 - (\xi - \frac{1}{6})R(x)\right]^{-1} ,
\label{eq0112}
\end{equation}
where $k^2 = \eta_{\mu\nu} k^\mu k^\nu$ and the coefficient $\overline f_j$ involve covariant terms formed from
the  Riemann tensor, its contractions and covariant derivatives \cite{parker2}.
It is important to note  that the expansion (\ref{eq0112}), although obtained using 
the Riemann normal coordinates \cite{Riemann}, is valid in a general coordinate system.
Three remarks are in order.
First, in the expansion of $\Delta$  increasingly high orders in curvature are compensated by increasingly high orders in $(k^2 - m^2)^{-1}$.
Second, any Feynman diagram in flat spacetime is substituted  in curved spacetime by an infinite set of diagrams. However, these diagrams have better UV convergences than the diagrams in flat Minkowski space.
Third, if one starts with a multiplicatively renormalizable theory in Minkowski spacetime, then in curved spacetime
 the total number of divergent diagrams, at any given loop order, is finite.
Hence, the locality of UV terms is secured \cite{shapiroQG}.

The Feynman propagator  (\ref{eq0012})  may be written as \cite{bunch}
\begin{equation}
\Delta (x,x')=(-g(x))^{-1/4}\bar{\Delta }(x,x'),
\label{eq1013}
\end{equation}
where the modified propagator $\bar{\Delta }$
satisfies 
\begin{equation}
(-\Box -m^2 -\xi R)\bar{\Delta }(x,x')=\delta(x-x') .
\label{eq2012}
\end{equation}
In Riemann normal coordinates 
the modified propagator takes the form
\begin{equation}
\bar{\Delta} (x,x')=\int\frac{d^{d}k}{(2\pi)^{d}}
 e^{-ik(x-x')}\left[ \frac{1}{k^2-m^2}+
\left(\xi-\frac{1}{6}\right)\frac{R}{(k^2-m^2)^2} + \ldots \right] .
\label{eq0013}
\end{equation}
where the ellipses denote the terms of higher curvature order.
Here, $x'$ is the origin of the Riemann normal coordinate system and $R$ is evaluated at the origin of these normal coordinates.

In the limit $x'\rightarrow x$,
the momentum integrals in (\ref{eq0013}) are divergent in 4 dimensions and must be regularized.
 Naively, one would expect that using a four-dimensional covariant cutoff automatically ensures the correct answer. 
Unfortunately, it is not so,  as Ossola and Sirlin \cite{ossola} demonstrated in flat spacetime. Even doing the regularization of zero point-energy of the scalar field one gets  wrong signs of the vacuum energy density vs.  pressure.
The most elegant way to deal with divergencies seems to be dimensional regularization (DR)\cite{veltman}. However, DR has its own obstacles. First, DR does not identify the quadratic divergencies which are important in any effective field theory.  Loosing the track of quadratic divergencies may corrupt the Wilson's renormalization group method. Next, since  DR is essentially a mass independent scheme, particle thresholds and particle decoupling is not properly described  and one is forced to put them in theory by {\it fiat}.
Ossola and Sirlin \cite{ossola} showed that the Pauli-Villars regularization, although  covariant, leads to unacceptable result: the vacuum energy density of a scalar field turns out to be negative. 
Besides, the quartic divergencies break the scale invariance of the free field theory in the massless limit.

Keeping all previously said in mind, we perform the calculations using the prescriptions of Cynolter and Lendvai \cite{cynolter}. Veltman has early noticed that a correct calculation of quadratic divergencies in $d = 2 - 2 ( \epsilon - 1)$ leads to a cutoff regularization based on the DR.
Doing correctly all conditions, preserving the symmetries, and using the prescription
\begin{equation}
l_{{\rm E}\mu}l_{{\rm E}\nu}\rightarrow\frac{1}{d}g_{\mu\nu}l_{\rm E}^{2}\label{eq:perd1}
\end{equation}
dictated by Lorenz invariance, with the parameter $d$  to be determined using the Euclidean four-dimensional momentum cutoff, a relation between the cutoff and the DR is obtained.
Matching different powers of $\Lambda_{\rm cut}$ and preserving the gauge invariance, the following
prescription is proposed \cite{cynolter}
\begin{eqnarray}
\frac{1}{d}\Lambda_{\rm cut}^{2} & \rightarrow & \frac{1}{2}\Lambda_{\rm cut}^{2},\label{eq:quad2}\\
\frac{1}{d}\ln\left(\frac{\Lambda_{\rm cut}^{2}+m^{2}}{m^{2}}\right) & \rightarrow & \frac{1}{4}\left(\ln\left(\frac{\Lambda_{\rm cut}^{2}+m^{2}}{m^{2}}\right)+\frac{1}{2}\right),\label{eq:log2}\\
\frac{1}{d} & \rightarrow & \frac{1}{4}\ \ \mathrm{for\ finite\ terms},
\label{eq:fin2}
\end{eqnarray}
which yields 
\begin{equation}
\int\frac{d^{d}k}{(2\pi)^{d}}\frac{1}{k^2-m^2}=-\frac{i}{(4\pi)^2} 
\left (\Lambda_{\rm cut}^2-m^2\ln\frac{\Lambda_{\rm cut}^2}{m^2} \right ) ,
\label{eq0015}
\end{equation}
\begin{equation}\label{eq0016}
\int\frac{d^{d}k}{(2\pi)^{d}}\frac{1}{(k^2-m^2)^2}=\frac{i}{(4\pi)^2} \left (\ln\frac{\Lambda_{\rm cut}^2}{m^2}-1 \right ) .
\end{equation}
Evaluating the integral over $m^2$ in (\ref{eq0014}) we find 
\begin{eqnarray}
& &
\Gamma_s=\frac{1}{2(4\pi)^2} \int d^4x \sqrt{-g}
\left[ \frac{m_s^4}{4} -m_s^2\Lambda_{\rm cut}^2 +m_s^2\left(\frac{m_s^2}{2}+\left(\xi -\frac{1}{6}\right) R\right)\ln\frac{\Lambda_{\rm cut}^2}{m_s^2}\right.
\nonumber \\
& &
+ c_1 \Lambda_{\rm cut}^4 +c_2 \Lambda_{\rm cut}^2R + \ldots \Big] ,
\label{eq0018}
\end{eqnarray}
where the ellipses denote the terms of higher adiabatic order.
The last two terms in square brackets are the $m_s^2$ independent ``constants'' of integration   where 
$c_1$ and $c_2$ are arbitrary dimensionless constants.

\subsection{Spinor fields}

The contribution of fermions to the effective action may be calculated in a
similar way. Taking a derivative with respect to $m_{\rm F}$ of the expression (\ref{eq0010}),
in which the action $S^{(2)}$ is replaced by
\begin{equation}
S^{(2)}_{\rm F}= \int d^4x \sqrt{-g}\,{\cal L}^{(2)}_{\rm F} ,
\label{eq104}
\end{equation}
with the Lagrangian  (\ref{eq0106}),
we find
\begin{equation}\label{eq0022}
\frac{\partial\Gamma_{\rm F}}{\partial m_{\rm F}}=\frac{i}{2}\int d^4x \sqrt{-g}\, {\rm tr}
\mathcal{S}(x,x) ,
\end{equation}
where the trace is taken over the spinor indices. 
The Feynman Green function $\mathcal{S}$ defined as
\begin{equation}\label{eq0023}
\mathcal{S}_{ab}(x,y)\equiv-i<T\Psi_a(x) \bar{\Psi}_b(y)> ,
\end{equation}
in the limit $y\rightarrow x$ may be expressed as \cite{parker}
\begin{equation}\label{eq0024}
\mathcal{S}(x,x)=\int\frac{d^{d}k}{(2\pi)^{d}}(-\gamma^{\mu}k_{\mu}+m_{\rm F}
-i\sqrt{2}\lambda\bar{\pi}\gamma_5)\mathcal{G}(k),
\end{equation}
where, to second adiabatic order,
\begin{equation}\label{eq0025}
\mathcal{G}(k) = \left [ 1+\frac{R}{12} \frac{\partial}{\partial m_{\rm F}^2}  \right ](k^2-m_{\rm F}^2)^{-1}.
\end{equation}
Using this we find
\begin{equation}
\frac{\partial \Gamma_{\rm F}}{\partial m_{\rm F}}=2i \int d^4x \sqrt{-g} \int \frac{d^nk}{(2\pi)^n} \left (\frac{m_{\rm F}}{k^2-m_{\rm F}^2}+\frac{m_{\rm F} R}{12(k^2-m_{\rm F}^2)^2} \right ) .
\label{eq0026}
\end{equation}
Finally, using (\ref{eq0015}) and (\ref{eq0016}) and integrating (\ref{eq0022}) over $m_{\rm F}$
we obtain 
\begin{eqnarray}
& &
\Gamma_{\rm F}=\frac{1}{(4\pi)^2} \int d^4x \sqrt{-g}
\left[m_{\rm F}^2\Lambda_{\rm cut}^2-\frac{m_{\rm F}^4}{4} -m_{\rm F}^2
\left(\frac{m_{\rm F}^2}{2}+\frac{R}{12}\right)\ln\frac{\Lambda_{\rm cut}^2}{m_{\rm F}^2} \right.
\nonumber \\
& &
  + c_3 \Lambda_{\rm cut}^4 +c_4 \Lambda_{\rm cut}^2 R  +\ldots \Big].
\label{eq0030}
\end{eqnarray}
Again, the last two terms in square brackets are the $m_{\rm F}$ independent ``constants'' of integration   where 
$c_3$ and $c_4$ are arbitrary dimensionless constants.

\subsection{All together}
\label{all}

Assembling the boson and fermion contributions, 
the resulting effective action for each species is
given by
 \begin{equation}
\Gamma=\Gamma_\sigma+\Gamma_\pi+\Gamma_{\rm F}
\label{eq0031}
\end{equation}
as a function of the background fields
$\bar{\sigma}$ and $\bar{\pi}$.
The background fields  are determined from 
the requirement that the vacuum minimizes  the potential (\ref{eq1105}).
One may easily verify that the function (\ref{eq1105}) has 
two minima: The first one is obviously the trivial minimum  at
$\bar{\pi}=0$, $\bar{\sigma}=0$ in which case all the 
effective masses are equal
$m_\sigma=m_\pi=m_{\rm F}=m$.
The second minimum occurs at 
$\bar{\pi}=0$, $\bar{\sigma}=-\sqrt{2}m/\lambda$.
In this case we find 
$m_\sigma=m_\pi=-m_{\rm F}=m$
and hence, this vacuum breaks the supersymmetry and yields a negative
fermion mass.
However, we immediately note that
the fermion contribution, being a function of $m_{\rm F}^2$, is the same for both solutions.
Hence,
both solutions yield a cancellation  of the
mass dependent flat spacetime contributions in (\ref{eq0031}), i.e., the mass dependent terms 
that do not vanish in the zero curvature limit in the fermion part $\Gamma_{\rm F}$ precisely cancel the corresponding terms in the boson parts $\Gamma_\sigma$ and $\Gamma_\pi$.

Next, we fix the unknown constants $c_1$ and $c_2$
in (\ref{eq0018}) and  $c_3$ and $c_4$ in (\ref{eq0030})
  from the following considerations \cite{sobreira}.
 Evaluating formally the functional integral in (\ref{eq0010}) over boson and fermion fields
 we may write the one loop effective action as
 \begin{equation}
\Gamma=2\Gamma_s+\Gamma_{\rm F}
=-2i\ln (\det D_s)^{-1/2}-i\ln \det D_F
=i \,{\rm tr}  \ln D_s -i\, {\rm tr} \ln D_{\rm F} ,
\label{eq0110}
\end{equation}
 where $D_s$ and $D_{\rm F}$ are the bilinear operators of the Lagrangians 
(\ref{eq0105}) and (\ref{eq0106}) (with $m_\pi=m_\sigma=m_{\rm F}=m$ and $\bar{\pi}=0$)
defined as
 \begin{equation}
D_s=   -\Box
-m^2-\xi R
\label{eq2105}
\end{equation}
and
\begin{equation}
D_{\rm F} = 
i\tilde{\gamma}^\mu\nabla_\mu- m .
\label{eq2106}
\end{equation}
Since the final result should depend on $m$ as a function of $m^2$ only, we can 
 simplify the trace of the fermion operator using
\begin{equation}
{\rm tr} \ln D_{\rm F}(m) = 
\frac{1}{2}{\rm tr} \ln D_{\rm F}(m) D_{\rm F}(-m)
={\rm tr}(-\Box
-\frac{1}{4} R- m^2),
\label{eq2107}
\end{equation}
where the factor 1/2 in front of the trace has been compensated by a factor of 2 
for the two degrees of freedom of the Majorana spinor.
Hence, the trace of the fermion operator is just
the trace of the scalar operator  with $\xi=1/4$.

Next, we use  (\ref{eq1012}) 
to invert the operator $D_s$ and calculate the trace using 
\begin{equation}
{\rm tr} \ln D_s = - {\rm tr} \ln \Delta  .
\label{eq2108}
\end{equation}
Working in normal coordinates, in this equation we may replace $\Delta$ by $\bar{\Delta}$
given by (\ref{eq0013}) which may be expanded as
\begin{equation}
\bar{\Delta}=\bar{\Delta}_0+\bar{\Delta}_1 R +\dots   ,
\label{eq2109}
\end{equation}
where $\Delta_0$ is the inverse of $\eta^{\mu\nu}\partial_\mu\partial_\nu -m^2$.
Hence, we have
\begin{equation}
{\rm tr} \ln D_s = - {\rm tr} \ln (\bar{\Delta}_0+\bar{\Delta}_1 R +\dots)
= -{\rm tr} \ln \bar{\Delta}_0 - {\rm tr}\, \bar{\Delta}_0^{-1} \bar{\Delta}_1 R +\dots \, .
\label{eq2110}
\end{equation}
The first term on the right-hand side of  this equation
is the flat spacetime contribution which cancels the similar term 
from the fermion part owing to (\ref{eq0110}) and (\ref{eq2107}).
Hence,  we only need to evaluate the last term in (\ref{eq2110})
\begin{equation}
{\rm tr}\, \bar{\Delta}_0^{-1} \bar{\Delta}_1 R=\int d^4x \sqrt{-g}\int d^4x'
\bar{\Delta}_0^{-1}(x,x') \bar{\Delta}_1(x',x) R ,
\label{eq2111}
\end{equation}
which, with a help of the Fourier transform yields
\begin{equation}
{\rm tr}\, \bar{\Delta}_0^{-1} \bar{\Delta}_1 R=\int d^4x \sqrt{-g}\int \frac{d^4k}{(2\pi)^4}
\left(\xi-\frac{1}{6}\right) \frac{R}{k^2-m^2} .
\label{eq2112}
\end{equation}
Using the regularized momentum integral (\ref{eq0015})
we find
\begin{equation}
2\Gamma_s=i {\rm tr} \ln D_s =  - i{\rm tr} \ln \bar{\Delta}_0
-\frac{1}{16\pi^2}\int d^4x \sqrt{-g} 
\left(\xi -\frac{1}{6}\right)\left(\Lambda_{\rm cut}^2- m^2 \ln\frac{\Lambda_{\rm cut}^2}{m^2}\right) R .
\label{eq1032}
\end{equation}
Comparing this result with (\ref{eq0018}) we find
\begin{equation}
c_2=\frac{1}{6}-\xi.
\label{eq2032}
\end{equation}
Using (\ref{eq2107})
the fermion part  is given by
\begin{equation}
\Gamma_{\rm F}= -i {\rm tr} \ln D_{\rm F} =- i {\rm tr} \ln D_s|_{\xi=1/4} ,
\label{eq3032}
\end{equation}
which yields
\begin{equation}
c_4=-c_2|_{\xi=1/4} =\frac{1}{12} .
\label{eq2033}
\end{equation}
Finally we find the total contribution to the effective action
\begin{equation}
\Gamma
=-\frac{1}{16\pi^2}\int d^4x \sqrt{-g}
N\tilde{\xi}\left(\Lambda_{\rm cut}^2- m^2 \ln\frac{\Lambda_{\rm cut}^2}{m^2}\right) R ,
\label{eq0032}
\end{equation}
where
\begin{equation}
\tilde{\xi} =\xi -\frac{1}{4} .
\label{eq2034}
\end{equation}
Owing to the cancellation of the first term on the right-hand side of (\ref{eq1032})
with a similar term in the fermion part  we obtain 
 $c_1+c_3=0$.
This secures that 
all the terms in (\ref{eq0030}) that do not vanish in the zero curvature limit,
precisely equal one half of the corresponding terms in the  scalar field expression (\ref{eq0018}),
with an opposite sign.
In this way, in the sum
(\ref{eq0031}) these terms precisely cancel 
 as they should,  
because, as demonstrated in \cite{bilic},
supersymmetry provides a  cancellation of all flat-spacetime contributions
 irrespective of the regularization method one uses.

\section{Energy Momentum Tensor}
\label{energy}

The vacuum expectation value of the energy momentum tensor is 
 derived from the effective action as
\begin{equation}\label{eq0034}
T^{\rm vac}_{\mu\nu}=\frac{2}{\sqrt{-g}}\frac{\delta \Gamma}{\delta g^{\mu\nu}} ,
\end{equation}
yielding
\begin{equation}
T^{\rm vac}_{\mu\nu}=-\frac{N}{8\pi^2}\tilde{\xi}\left( \Lambda_{\rm cut}^2 -
m^2
\ln\frac{\Lambda_{\rm cut}^2}{m^2}  \right)
\left(R_{\mu\nu}-\frac{1}{2} g_{\mu\nu}R\right).
\label{eq0035}
\end{equation}
First, we note that  we generally  do not recover the vacuum form of the energy momentum tensor
\begin{equation}
T_{\mu\nu}=\rho g_{\mu\nu}.
\label{eq328}
\end{equation}
Clearly, the energy momentum tensor (\ref{eq0035}) cannot be put in the form (\ref{eq328})
unless the metric satisfies 
\begin{equation}
R_{\mu\nu}
 \propto g_{\mu\nu} R .
\label{eq0328}
\end{equation}
Among homogeneous geometries, this equation holds  only for Minkowski and de Sitter spacetimes.

Now, we consider the case when the starting background is provided by pure gravity 
with the cosmological constant  ${\Lambda}$. 
Adding (\ref{eq0032}) to the Einstein-Hilbert action we obtain the total action
\begin{equation}
S
=\frac{1}{16\pi G}\int d^4x \sqrt{-g} \left[R-\frac{GN\tilde{\xi}}{\pi}
\left(\Lambda_{\rm cut}^2-m^2 \ln\frac{\Lambda_{\rm cut}^2}{m^2}\right)
R 
- 2 {\Lambda}\right]  ,
\label{eq0132}
\end{equation}
which may be recast in the  standard Einstein-Hilbert form
\begin{equation}
S
=\frac{1}{16\pi G_{\rm eff}}\int d^4x \sqrt{-g} \left(
R 
- 2 {\Lambda}_{\rm eff}\right). 
\label{eq0133}
\end{equation}
Here we have introduced the effective Newton and cosmological constants 
given by
\begin{equation}
\frac{G_{\rm eff}}{G}=\frac{\Lambda_{\rm eff}}{{\Lambda}}
=\lambda ,
\label{eq0134}
\end{equation}
where
\begin{equation}
\lambda
=\left[1-\frac{GN\tilde{\xi}}{\pi}\left(\Lambda_{\rm cut}^2-m^2 \ln\frac{\Lambda_{\rm cut}^2}{m^2}
\right)\right]^{-1} .
\label{eq0234}
\end{equation}
 More specifically, if we demand homogeneity and isotropy
the starting background geometry is de Sitter
with the expansion parameter $H=\sqrt{{\Lambda}/3}$.
Then, the resulting effective background is also  de Sitter with the effective expansion parameter
 $H_{\rm eff}=\sqrt{{\Lambda}_{\rm eff}/3}$.

Consider next a more general case of gravity
with matter or dark energy described by the energy momentum tensor $T_{\mu\nu}$.
The total energy momentum tensor is the sum of $T_{\mu\nu}$ 
and  $T^{\rm vac}_{\mu\nu}$ given by (\ref{eq0035}). 
The Einstein field equations then read
\begin{equation}
\left(R_{\mu\nu}-\frac{1}{2} g_{\mu\nu}R\right)=\frac{GN\tilde{\xi}}{\pi}\left(\Lambda_{\rm cut}^2 
-m^2\ln\frac{\Lambda_{\rm cut}^2}{m^2}  \right)
\left(R_{\mu\nu}-\frac{1}{2} g_{\mu\nu}R\right) -8\pi G T_{\mu\nu} ,
\label{eq0135}
\end{equation}
which may be written as
\begin{equation}
\left(R_{\mu\nu}-\frac{1}{2} g_{\mu\nu}R\right)= -8\pi G \lambda T_{\mu\nu}.
\label{eq0136}
\end{equation}
This equation may be interpreted as the standard Einstein equations in which
either the Newton constant $G$  is renormalized as in (\ref{eq0134}) or
the energy momentum tensor is replaced by the effective one
\begin{equation}
T^{\rm eff}_{\mu\nu} = \lambda T_{\mu\nu}.
\label{eq0137}
\end{equation}
Assuming a general perfect fluid form 
\begin{equation}
T_{\mu\nu}=
(\rho+p)u_\mu u_\nu -p g_{\mu\nu} ,
\label{eq616}
\end{equation}
 the energy density and pressure  
in comoving coordinates are
\begin{equation}
\rho =<{T^0}_0> ,
\label{eq227}
\end{equation}
\begin{equation}
p =\frac{1}{3}<T_0^0-{T^\mu}_\mu>.
\label{eq327}
\end{equation}
Obviously,  equations (\ref{eq616})-(\ref{eq327}) also apply to $T^{\rm eff}_{\mu\nu}$ 
defined in (\ref{eq0137}), with $\rho$ and $p$ replaced by $\rho_{\rm eff}$ and $p_{\rm eff}$, 
respectively.
 Hence, if the energy density and pressure satisfy an equation of state of a simple form
$p = w \rho$,
then the effective equation of state
is of the same form $p_{\rm eff} = w \rho_{\rm eff}$.
In particular, if the background satisfies the vacuum equation of state 
$p_{\rm vac} = - \rho_{\rm vac}$, the effective equation of state also describes vacuum.
However, if the equation of state were more involved, e.g., in the form of a general function
$p=p(\rho)$ then the effective equation of state would be deformed, 
$p_{\rm eff}=\lambda p(\rho_{\rm eff}/\lambda)$.

In order to make comparison with a similar calculations in which a three-dimensional momentum cutoff 
has been used \cite{bilic2}
(see also \cite{maggiore} and 
\cite{sloth} for massless scalars) we specify our result to a spatially flat FRW spacetime
and $\xi=0$.
 From (\ref{eq0035}), we find
the leading  contribution to  the density and pressure 
as a function of the expansion rate and acceleration 
\begin{equation}
\rho_{\rm vac}=-\frac{3N\Lambda_{\rm cut}}{32\pi^2}\frac{\dot{a}^2}{a^2}\, ,
\label{eq0236}
\end{equation}
\begin{equation}
p_{\rm vac}=
\frac{N\Lambda_{\rm cut}}{32\pi^2}\left(\frac{\dot{a}^2}{a^2}+2\frac{\ddot{a}}{a}\right).
\label{eq0336}
\end{equation}
Obviously, the contribution to the energy density is negative
and to the pressure is positive for an accelerated expansion.
 In contrast, a three-dimensional momentum cutoff 
 regularization 
 yields 
\cite{bilic2}
\begin{equation}
\rho=\frac{3N\Lambda_{\rm cut}}{24\pi^2}\frac{\dot{a}^2}{a^2}\, ,
\label{eq236}
\end{equation}
\begin{equation}
p=
\frac{N\Lambda_{\rm cut}}{24\pi^2}\left(\frac{\dot{a}^2}{a^2}-2\frac{\ddot{a}}{a}\right).
\label{eq336}
\end{equation}
Here, the density is positive and the
pressure is negative if $\dot{a}^2<2a\ddot{a}$.
For example, for a de Sitter expansion one finds  $p= -\rho/3$
whereas in the present paper $p_{\rm vac}= -\rho_{\rm vac}$.
Hence, apart from an irrelevant numerical factor (which may be absorbed in the cutoff)
and the relative sign between the expansion and acceleration terms in the pressure,
we  disagree in the overall sign but we agree in the magnitude of the energy density.
In both cases a UV cutoff of the order $m_{\rm Pl}/\sqrt{N}$ yields a magnitude of 
the leading term in the energy density  of the order
$H^2 m_{\rm Pl}^2$, where $H=\dot{a}/a$. Hence, if we identify the  expansion parameter $H$ with the Hubble parameter today, the contribution to CC is of the 
phenomenologically acceptable order of magnitude and no fine tuning is needed.

Clearly, the above comparison between results obtained with a three-dimensional momentum cutoff and with the present covariant   cutoff regularization  does not concern renormalized quantities. Instead, our aim is to compare the results of the two approaches to an effective field theory with a physical cutoff \cite{lepage},
 and, hence, we do not have bare and renormalized $\Lambda$ and $G$, as in \cite{maggiore2,maggiore3}
(see also \cite{babic}).
This approach is in spirit similar to \cite{weinberg2} 
 where a discrepancy by a factor of about $10^{120}$ 
was obtained  in  an effective theoretical  estimate of the  vacuum energy density
with respect to the value compatible with astrophysical and cosmological observations.

Although our aim is not to fit observational data, it is important to note  that one cannot have a
pure $H^2 m_{\rm Pl}^2$ for the vacuum energy without additional contributions.
At the phenomenological level,
it has been already pointed out that  only after including an additive term  one
can get full compatibility with the expansion and structure formation
data \cite{sola2,sola4}.

\section{Conclusion}
\label{conclude}

We have calculated the contribution of supersymmetric fields to vacuum energy 
in a general curved geometry.
In addition to supersymmetric fields we have assumed existence of a fluid obeying the equation of state
$p=p(\rho)$.
Unlike in flat spacetime, the vacuum fluctuations turn out to be nonzero depending 
on background metric. Combining effects of both
the background fluid and vacuum fluctuations of the supersymmetric field
in a self-consistent way, we have
found the effective equation of state.

In an expanding FRW universe with $H=\dot{a}/a$, we have found that 
the leading term in the energy density of vacuum fluctuations is   negative and,
if we impose a UV cutoff of the order $m_{\rm Pl}$ 
its magnitude is of the order
$H^2 m_{\rm Pl}^2$. Hence, the vacuum fluctuations of SUSY fields  provide
a phenomenologically acceptable contribution to the cosmological constant
and no fine tuning is needed.
The negative sign of the energy density indicates that the vacuum field fluctuations 
cannot account for CC alone, hence, one additional positive contribution, 
either in the form of CC or a more general form of dark energy,
is needed.

\subsection*{Acknowledgments}
We are grateful to Ilya Shapiro and Joan Sola for useful discussions and comments.
This work was supported by the Ministry of Science,
Education and Sport
of the Republic of Croatia under Contract No. 098-0982930-2864.

\end{document}